\documentclass[prb,preprint,showpacs,floatfix,longbibliography]{revtex4-2}

\usepackage{hyperref}
\usepackage{graphicx}
\usepackage{array}
\usepackage{amsmath,amsfonts,amssymb}
\usepackage{color}
\usepackage{calc}
\usepackage[normalem]{ulem}
\usepackage{svg}
\usepackage{natbib}
\usepackage{bibentry}
\usepackage{braket}
\usepackage[english]{babel}

\newcommand{\Figref}[1]{Fig.~\ref{#1}}

\begin{document}

\title{Topological Frenkel excitons in 2D molecular crystals}

\author
{Quentin Stein$^{1}$, Sidhartha Nayak$^{2}$, Song Jiang$^{1}$, Jakob Kuhlke$^{3}$, Michelangelo Romeo$^{1}$, Thomas Frederiksen$^{4,5}$, Fabrice Scheurer$^{1}$, Katharina Kaiser$^{1,3}$, Tom\'{a}\v{s} Neuman$^{2}$, Guillaume Schull$^{1}$\\
\normalsize{$^1$Universit\'e de Strasbourg, CNRS, IPCMS, UMR 7504, F-67000 Strasbourg, France.} \\
\normalsize{$^{2}$Institute of Physics, Czech Academy of Sciences}\\
\normalsize{Cukrovarnick\'{a} 10, Prague, 16200, Czech Republic}\\
\normalsize{$^{3}$IV. Physical Institute -- Solids and Nanostructures,\unpenalty~Georg-August-Universit\"{a}t G\"{o}ttingen, G\"{o}ttingen,\unpenalty~37077,\unpenalty~Germany}\\
\normalsize{$^{4}$Donostia International Physics Center,\unpenalty~Donostia-San Sebasti\'{a}n 20018,\unpenalty~Spain}\\
\normalsize{$^{5}$Ikerbasque, Basque Foundation for Science, E-48013 Bilbao, Spain}\\}

\begin{abstract}
Collective molecular excitations govern energy transport in light-harvesting and optoelectronic materials. Theory predicts that excitonic band topology can generate edge-confined states, offering a route to direct energy flow in molecular systems, yet experimental evidence is lacking. Here we use scanning tunnelling microscopy-induced luminescence to image Frenkel excitons in two-dimensional zinc phthalocyanine molecular crystals with nanometer resolution. By varying cluster size, we observe the transition from discrete quantum-confined excitations to collective excitonic bands and directly resolve their real-space wavefunctions. We identify an excitonic state localized at the crystal perimeter and show, through comparison with microscopic theory, that it originates from the momentum-resolved non-trivial topology of the bulk excitonic bands. Our results establish molecular crystals as a platform for engineering topological excitonic states and provide a strategy for controlling energy transport through excitonic band engineering.        
\end{abstract}

\maketitle
\noindent
Controlling energy flow in molecular systems is a central challenge in photophysics, with implications for light harvesting, molecular optoelectronics, organic photovoltaics, and quantum information technologies \cite{Scholes2003,Agranovich2008}. At the molecular scale, energy transfer 
is dominated by dipole-dipole coupling between electronically excited molecules. In the coherent limit, these interactions hybridize local excitations into collective states extending over multiple molecular units \cite{Davydov1962,Agranovich2008,Wurthner2011}. In ordered two-dimensional (2D) molecular assemblies, such coupling gives rise to Frenkel excitons forming continuous energy bands \cite{Eisfeld2017,Juergensen2023,Kim2023,Hasan2025}. The associated band structure governs how excitations spatially distribute and propagate, making the engineering of excitonic bands a promising route to control energy flow in molecular architectures\cite{Knoester2006,Nematiaram2021}.
Recent advances in condensed matter physics have revealed that topology often accounts for wave localization and transport, and can thus serve as a guiding principle for designing robust quantum states of matter. In two-dimensional systems, band topology can give rise to boundary states confined to edges or interfaces providing propagating channels for particles and quasiparticles.
These concepts are well established for electronic, photonic and phononic systems \cite{Hasan2010,Qi2011}, and have even been predicted for Frenkel excitons in molecular lattices featuring non-trivial Chern topology \cite{Zhou2014}. In this latter case, the topology of bulk excitonic bands implies the emergence of localized edge excitons, offering a strategy to steer and confine energy flow at the nanoscale.
            
Whether a topologically confined edge state can be realized experimentally for excitons in 2D molecular assemblies remains an open question. Addressing this challenge requires access to the spatial structure of excitonic eigenstates. Recent developments in scanning tunnelling microscopy-induced luminescence (STML) have enabled optical imaging of molecular excitations with sub-molecular spatial resolution, providing direct access to the spatial structure of excitonic states \cite{Imada2016,Zhang2016,Doppagne2017,Roslawska2022,Dolezal2021b,Luo2026,Jing2026} yet real-space visualization of Frenkel excitons and their band states in nanoscale 2D assemblies has so far remained out of experimental reach.

Here, we report the experimental and theoretical characterization of Frenkel excitons in two-dimensional crystals of zinc phthalocyanine (ZnPc) molecules using STML. 
Small clusters behave as molecular quantum dots with discrete optical resonances reflecting strong spatial confinement. As size increases, this confinement is lifted and a coherent excitonic band emerges, whose real-space wavefunctions are directly resolved in STML maps.
Strikingly, for specific edge geometries of the molecular crystals, the momentum-resolved topology of these bands features a non-trivial Zak phase \cite{Zak1989,Delplace2011}, giving rise to edge-confined states. We further investigate the impact of intentionally introduced molecular vacancies on the resilience of the excitonic edge-bands.
Our results provide real-space visualization of bulk and edge-localized Frenkel excitons in a two-dimensional molecular cluster and establish engineering of topology as a route to achieve resilient and directional energy transport in coherently coupled molecular nanostructures.

\begin{figure}
  \includegraphics[width=\linewidth]{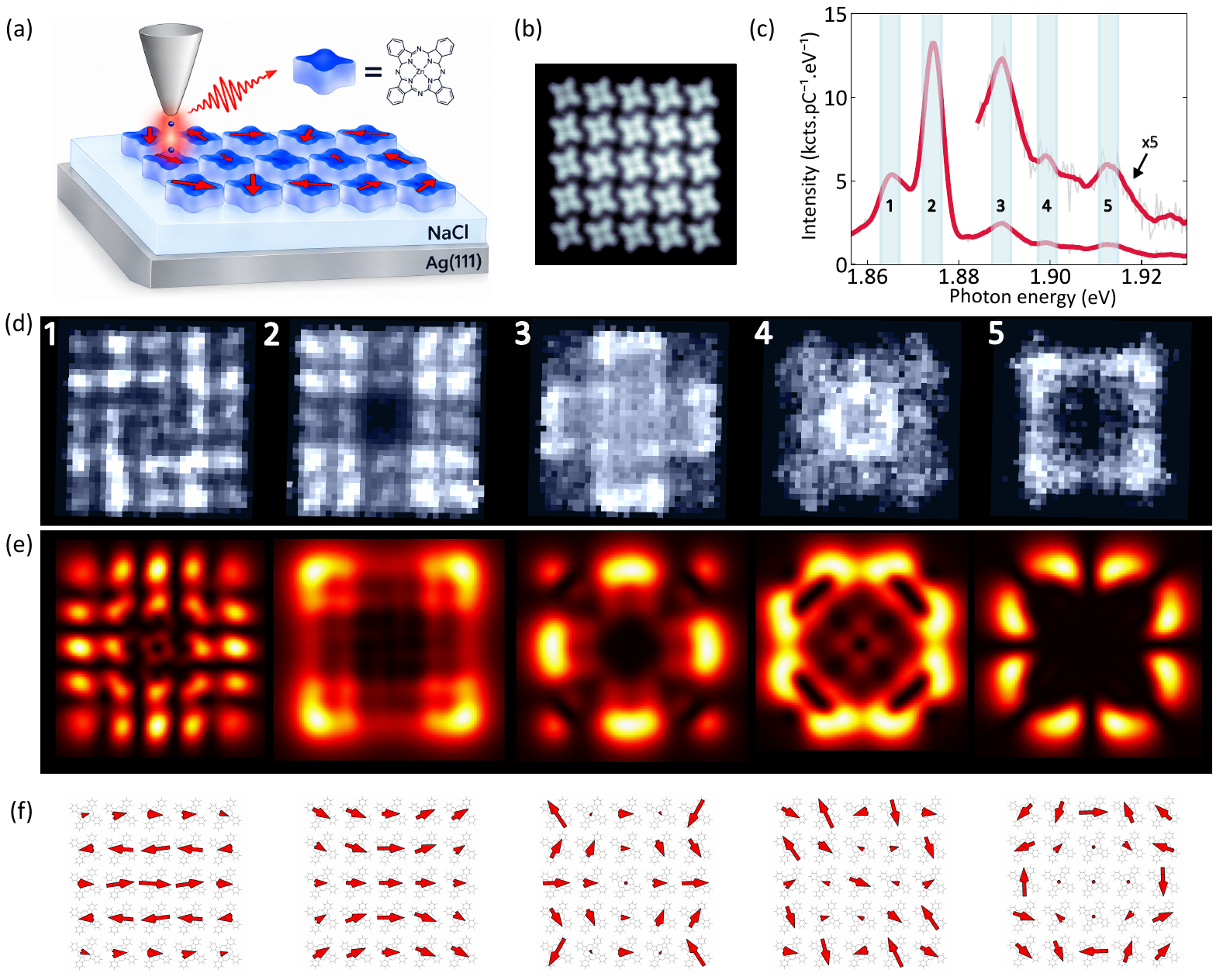}
  \caption{\label{figure1}
Frenkel excitons in self-assembled 2D ZnPc clusters. 
(a) Schematic of the experiment: Frenkel excitons confined in ZnPc molecular clusters on 3ML-NaCl/Ag(111) are probed by STML. 
(b) STM topography (8.3 $\times$ 8.3 nm$^2$, tunneling current $I$ = 5 pA, sample bias $V$ = 500 mV) of a 5$\times$5 cluster.
(c) STML spectrum obtained by averaging 1225 STML spectra recorded on the cluster over a 35 $\times$ 35 pixel grid. Each spectrum was acquired during 35 s at $V$ = - 2.7 V with a current ranging from 0.2 pA to 375 pA. The average spectrum is showing five emission features between 1.86 eV and 1.93 eV.
(d) Hyperspectral STML maps (8.3 $\times$ 8.3 nm$^2$) obtained by integrating the emission intensity for the five spectral features identified in (c). 
(e) Simulated plasmon-exciton coupling maps (see SI for details) of selected collective excitonic modes (modes 1-5). 
(f) Corresponding collective dipole configurations (one representative of each doubly degenerate pair for modes 1, 2, 3 and 4), illustrating the underlying phase relationships governing bright, dark, and edge-confined excitonic states. The length of the red arrows is proportional to the amplitude of the dipole.}
\end{figure}

Figure\,\ref{figure1}a schematically illustrates the experiment, in which the fluorescence properties of an engineered ZnPc molecular cluster adsorbed on 3ML-NaCl/Ag(111) are probed by scanning tunneling microscopy-induced luminescence (STML). An STM image of a 5 $\times$ 5 ZnPc cluster is shown in \Figref{figure1}b. The procedure used to assemble the two-dimensional molecular clusters is described in the Supplementary Information (SI). By adjusting the tunneling conditions during image acquisition (see SI), the molecular positions can be determined with respect to the NaCl lattice with sub-molecular precision. This analysis reveals a commensurate organization of the clusters on NaCl, resulting in highly regular structures with nearly identical intermolecular distances.

The STML spectrum displayed in \Figref{figure1}c was obtained by averaging all 1225 spectra acquired on a regular 35 $\times$ 35 grid over the cluster (all individual spectra are provided in SI as a concatenated heat map). The resulting spectrum reveals five distinct features between 1.86 and 1.93 eV. This energy range is characteristic of the fluorescence of ZnPc monomers and small aggregates\cite{Doppagne2017}, suggesting that these peaks originate from Coulomb coupling between chromophores. While spectra recorded at different tip positions exhibit pronounced variations in the relative peak intensities (see SI), the number of features remains fixed at five and their energies vary only marginally.

To visualize the spatial distribution of these spectral features, we constructed hyperspectral fluorescence maps (\Figref{figure1}d). STML spectra were acquired at each tip position across the cluster, and the fluorescence intensity maps were obtained by integrating the intensity of the emitted light within energy windows corresponding to the five spectral features. Throughout the measurements, the tip-sample distance was kept constant (open-feedback mode) to avoid artifacts arising from variations in plasmon-exciton coupling. The maps were normalized by the local tunneling current to compensate for position-dependent excitation probabilities\cite{roslawska2026}. The resulting experimental maps reveal five distinct spatial patterns.
To interpret those, we developed a Frenkel-exciton model describing coherent excitonic coupling within the molecular cluster. We consider that each molecular monomer hosts two degenerate singlet excited states, $S_\alpha$ ($\alpha\in\{x,y\}$), that give rise to two degenerate transitions $Q_\alpha$ to the ground state. Each transition is associated with its corresponding transition charge density $\rho_\alpha$ carrying a dipole moment ${\bf p}_\alpha$. The coupling between transition $Q_\alpha$ in monomer $m$ at position ${\bf R}_m$ and transition $Q_\beta$ of monomer $n$ at position ${\bf R}_n$ is then $V_{mn}^{\alpha\beta}=\iint {\rho_\alpha  ({\bf r}-{\bf R}_m)\rho_\beta  ({\bf r}'-{\bf R}_n)}/{|{\bf r}-{\bf r}'|}{\rm d}^3{\bf r}{\rm d}^3{\bf r}'$ ($\alpha, \beta\in \{x,y\}$), yielding a Hamiltonian
\begin{align}
    \hat{H}=\sum_{\substack{m\neq n \\ \alpha\beta}}V_{mn}^{\alpha\beta}\vert n,\beta\rangle\langle m,\alpha\vert,\label{eq:hamiltonian}
\end{align}
where $\vert n,\beta\rangle$ is the excited state $\vert {S}_\beta  \rangle$ of monomer $n$, considering all other molecules in their ground state; the states are thus spanning the single-excitation subspace of the crystal.
The molecular transition densities $\rho_{\alpha}({\bf r})$ were calculated at the time-dependent density-functional theory (TDDFT) level and approximated by atom-centered point charges, as described in [\onlinecite{Nayak2026}]. The coupling between the plasmonic cavity of the STM junction and the collective excitonic state $j$ was evaluated using $G_j({\bf r}_{\rm tip})=\sum_{\alpha,i}c_{\alpha, i}^{(j)}g_{\alpha, i}({\bf r}_{\rm tip})$. Here, $c_{\alpha, i}^{(j)}$ are the coefficients of the collective eigenstate $j$ obtained by diagonalizing $\hat{H}$ [Eq.\,\eqref{eq:hamiltonian}], and $g_{\alpha, i}({\bf r}_{\rm tip})=\int \rho_{\alpha}({\bf r}-{\bf R}_i)\phi_{\rm pl}({\bf r}-{\bf r}_{\rm tip})\,{\rm d}^3{\bf r}$ is the coupling between the plasmon, represented by the quasistatic potential $\phi_{\rm pl}({\bf r}-{\bf r}_{\rm tip})$ at tip position ${\bf r}_{\rm tip}$, and the transition density of the molecule at ${\bf R}_i$. See SI and reference \onlinecite{Nayak2026} for details. The plasmon-exciton coupling map of mode $j$ then corresponds to $|G_j({\bf r}_{\rm tip})|^2$. The model generates $2\times N_{\rm mol}$ collective modes arising from coherent coupling among the $2\times N_{\rm mol}$ molecular transition dipoles present in the cluster of $N_{\rm mol}$ molecules (two per molecule). However, only a small subset of these modes couples efficiently to the plasmonic cavity and therefore contributes significantly to the measured luminescence (see SI). In \Figref{figure1}e we display simulated plasmon-exciton coupling maps for a selection of modes that reproduce the experimental STML maps (see SI for details about mode selection).
The corresponding collective dipole distributions are presented in \Figref{figure1}f. For convenience, only one representative dipole pattern is shown for the doubly degenerate modes 1, 2, 3 and 4. Mode 1 exhibits an alternating arrangement of dipoles in a given column, while being in-phase along the rows. The overall net dipole of this mode is negligible, characterizing a dark collective excitonic state. Despite its nominally forbidden optical character, this mode remains visible experimentally. This behavior results from the combined action of two mechanisms: First, rapid internal relaxation funnels population into the lowest-energy excitonic state. Second, coupling to the plasmonic cavity partially relaxes the optical selection rules, enabling radiative recombination from this otherwise dark mode \cite{Zhang2016,Neuman2018, roslawska2026}.
In contrast, mode 2 is characterized by an in-phase oscillation of all molecular dipoles across the cluster. This superradiant configuration is associated with a large transition dipole moment. Consequently, mode 2 produces an intense spectral feature despite the presence of efficient non-radiative relaxation pathways toward mode 1. The combination of its large oscillator strength and plasmon-enhanced radiative decay allows this bright state to remain prominently visible. The same mechanisms account for the observation of modes 3, 4, and 5.
Compared with mode 2, mode 4 exhibits a more complex spatial pattern, corresponding to a higher-order excitonic standing-wave mode supported by the finite-sized cluster.
Mode 5 stands out in that it localizes almost exclusively at the cluster periphery. This is reproduced by the simulations, which reveal a nearly vanishing net transition dipole moment for the central molecules, confirming the edge-confined character of this collective excitonic state. A similar pattern is observed for mode 3, which is also localized dominantly on the edge, but to a lesser extent. 

Until now, excitonic states of large 2D molecular aggregates have primarily been investigated using ensemble-averaged far-field techniques, and their spatial distributions could only be inferred indirectly from theoretical models\cite{Eisfeld2017,Juergensen2023,Kim2023,Hasan2025}. Our approach enables the \textit{direct} imaging of individual clusters and provides real-space visualization of collective excitonic modes and their spatial structure. 

\begin{figure}
  \includegraphics[width=\linewidth]{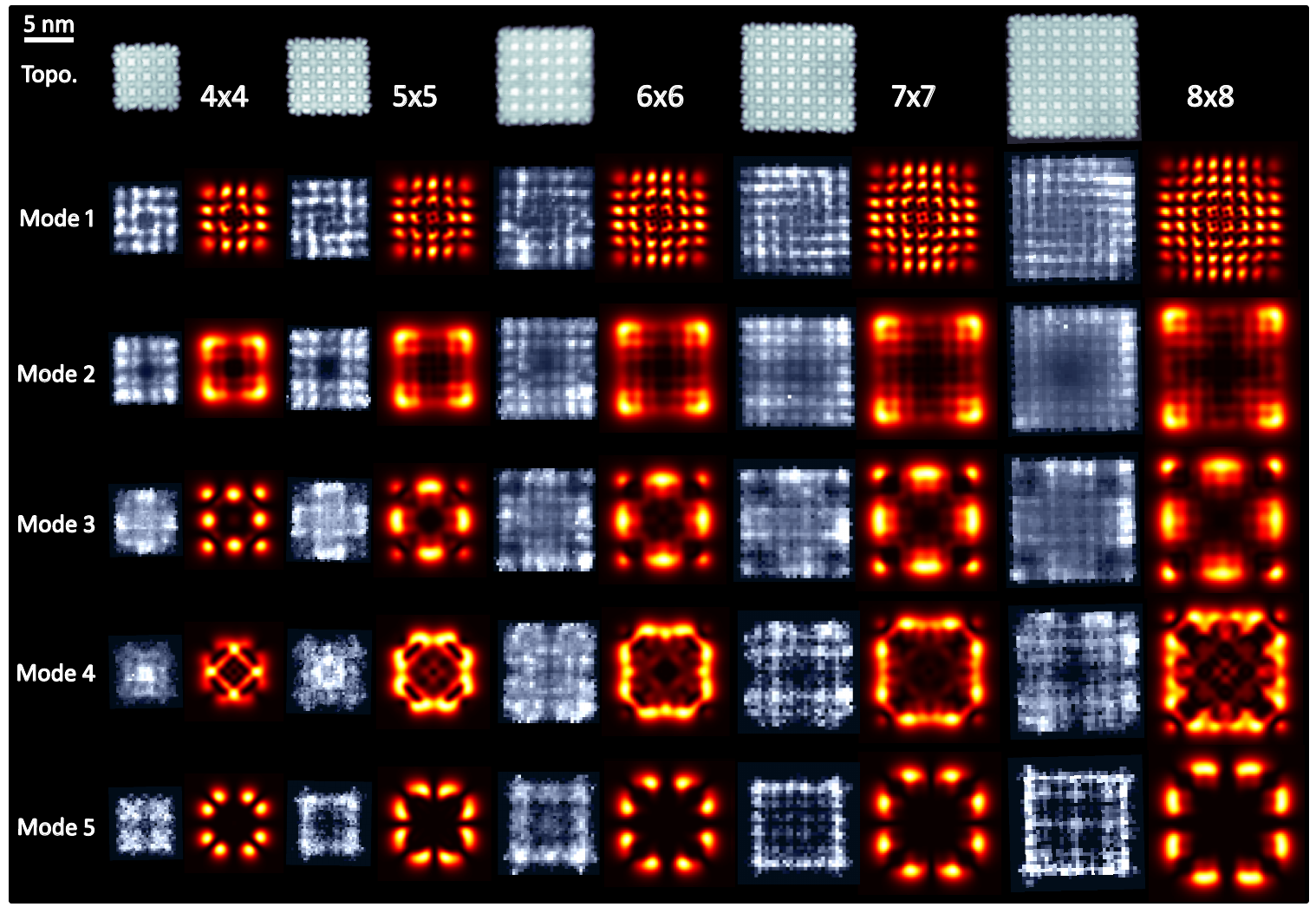}
\caption{\label{figure2}
Size evolution and emergence of excitonic band structure in finite 2D ZnPc clusters. 
Experimental photon maps (blue) and corresponding simulated plasmon-exciton coupling maps (red) for the main collective excitonic spectral features for $N \times N$ clusters with $4 \leq N \leq 8$ (see Supplementary information for details of the mode selection). The top row shows the corresponding experimental topography images.
The experimental STML maps ($V$ = - 2.7 V, with a current ranging from 0.2 pA to 375 pA) were obtained by integrating the emission intensity over the main features identified in the spectra.}
\end{figure}

In \Figref{figure2}, we analyze the evolution of these modes as a function of cluster size, presenting both experimental fluorescence maps and simulated plasmon-exciton coupling maps for $N \times N$ clusters with $4 \leq N \leq 8$. Remarkably, for $N \geq 5$, both the number of states and their spatial distributions remain highly similar. In particular, mode 5 stays confined to the cluster edge even for the largest sizes.
For $N = 4$, the first four modes already reproduce the spatial structure observed in larger clusters, whereas mode 5 begins to deviate. For $N = 3$ (See SI and Ref.~\onlinecite{Jing2026}), the spatial profiles are no longer comparable to those of larger clusters, with the exception of the two lowest-energy modes. 
These qualitative changes originate 
from the fact that, for these small clusters, a ''bulk'' can no longer be defined. In this sense, $N = 3$ clusters should be regarded as excitonic quantum dots, whereas for $N \geq 4$,  the clusters gradually develop characteristics of two-dimensional periodic nanocrystals and an associated Frenkel-exciton band structure.

\begin{figure}
  \includegraphics[width=0.95\linewidth]{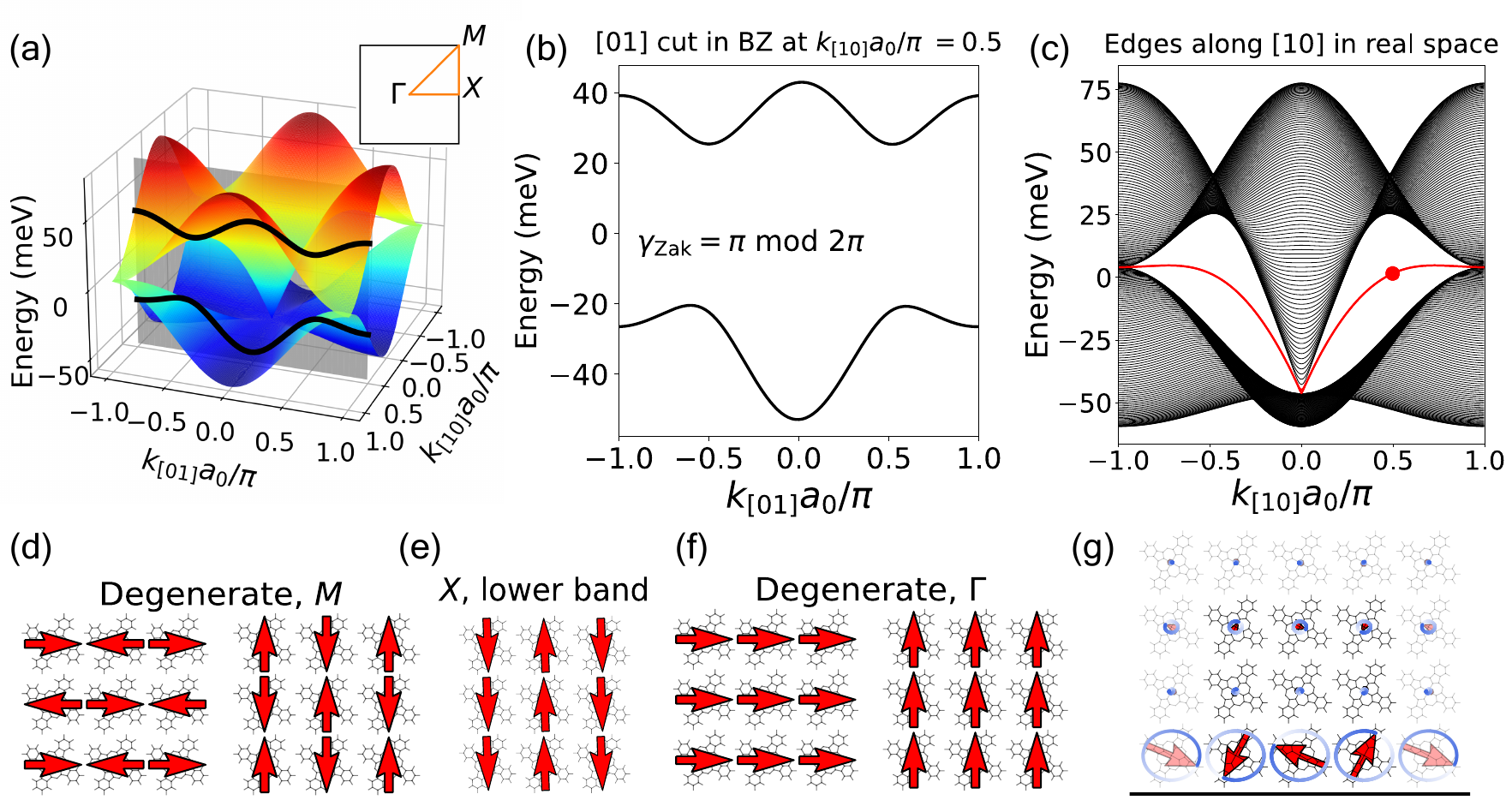} 
  \caption{\label{fig:theory}
Band structure of 2D and 1D excitons. (a) Band structure of the 2D aggregate. 
(b) Example of a cross section through the 2D band structure for fixed $k_{[10]}=\pi/(2a_0)$, which takes the form of a 1D two-band structure with a clearly defined gap. This allows us to evaluate the Zak phase for each $k_{[10]}$. (c) 1D band structure after introducing a [10] edge. (d)-(f) Real-space distribution of the excitonic wave functions on a $3\times 3$ sample of the infinite crystal evaluated at (d) M, (e) X, and (f) $\Gamma$. The excitons are visualized using transition dipoles of each monomer shown as red arrows. (g) Real-space distribution of the edge exciton marked by the red dot in (c). The ellipses represent the polarization of the individual molecules. For phase information we also include arrows indicating the orientation of the transition dipole moment at time $t=0$. 
}
\end{figure}

The emergence of excitonic bands for $N \geq 4$ is well captured by simulations of an infinite 2D crystal ($N \to \infty$). We show the theoretically calculated energy bands of the infinite crystal with lattice parameter $a_0=1.45$\,nm in Fig.\,\ref{fig:theory}a. We theoretically analyze this system in detail in Ref.~\onlinecite{Nayak2026}, where we extend this approach to semi-infinite 1D periodic geometries, i.e. introduce edges. Different cutting angles, i.e. edge geometries, have been investigated. In Fig.\,\ref{fig:theory}c we reproduce the 1D projected band structure for the edge cut along [10] corresponding to the one observed in the experimental clusters.
Here, the black dispersion lines correspond to the bulk bands projected into 1D and the red line indicates edge states emerging in the projected bulk band gap. 
As shown in Fig.\,\ref{fig:theory}g, this band corresponds to edge-confined states. In Ref.~\onlinecite{Nayak2026} we show that their origin can be traced back to the momentum-resolved topology of the 2D exciton. In other words, for each momentum $k_{[10]}$ along the edge, we define an effective 1D Hamiltonian in the $k_{[01]}$ direction (as indicated in Fig.\,\ref{fig:theory}a). This Hamiltonian, whose band structure can be obtained as a cut through the 2D bands shown in Fig.\,\ref{fig:theory}b, has a similar algebraic structure to that of the well-known Su-Schrieffer-Heeger model \cite{Su1979, Su1980}. Here, the Zak phase is known to indicate the presence or absence of edge states depending on how the infinite system is terminated. 
More concretely, the edge states emerge due to a non-trivial  $\gamma=\pi$ Zak phase  \cite{Berry1984, Zak1989, Xiao2010, Delplace2011}. However, in our case, the two-element basis of the unit cell is formed by a pair of degenerate excitations rather than an atomic dimer, thus uniquely defining both the physical termination and the choice of unit cell. 
As a result, the sense of their generally elliptical polarization is linked to the direction of propagation. Moreover, the confinement of these edge states is proportional to the size of the projected band gap at a given value of $k_{[10]}$ along the edge (Fig.\,\ref{fig:theory}c), providing some protection against scattering to the bulk states. The real-space polarization texture of an edge state for $k_{[10]}=\frac{\pi}{2a_0}$ is shown in \Figref{fig:theory}g. Here the arrows indicate instantaneous polarization and the blue ellipses indicate the polarization state. The sense of rotation of the polarization vector is visualized by the ellipse shading.   
In finite-sized clusters, these edge states form standing waves of counter-propagating excitons that are eventually imaged in the experiment (mode 5). These interfering waves give rise to a texture of linearly polarized monomer excitations whose angle of tilt varies along the edge. The time-dependent rotation of the polarization vectors thus turns into their spatial dependence along the edge. 

In the experiment, the peaks observed in the light spectrum (\Figref{figure1}c) correspond to the exciton modes of lower energy. The band structure in \Figref{fig:theory}a reveals that the excitons of lowest energy belong to the lower band of the bulk excitons. We therefore expect to identify the polarization textures found in the finite-sized clusters predominantly with those of the confined excitonic modes of the lower exciton branch. We show these textures corresponding to the lower branch at three significant points of the Brillouin Zone, X, $\Gamma$, and M, in \Figref{fig:theory}d-f. These textures can be assigned to peaks 1, 2, and 4 in \Figref{figure1}c. The remaining peaks 3 and 5, however, do not correspond to any of the bulk states, but can be assigned to the edge excitons of different $|k_{[10]}|$ forming standing waves confined to the square.  

A direct analogy can be drawn with graphene nanoribbons, where electronic states originating from momentum-resolved topology of 2D graphene are localized along zigzag edges\cite{Delplace2011, Niimi2006, Kobayashi2005, Tao2011, Wang2016a}. Similarly, in finite clusters with appropriate edge termination, this in-gap band manifests as a mode localized at the crystal periphery, providing a natural explanation for mode 5 observed in our system. We further note that momentum-locking of circular polarization, encoded in the oscillation of the individual transition dipoles, may suppress backscattering, which could contribute to the robustness of these edge states.

\begin{figure}
  \includegraphics[width=0.6\linewidth]{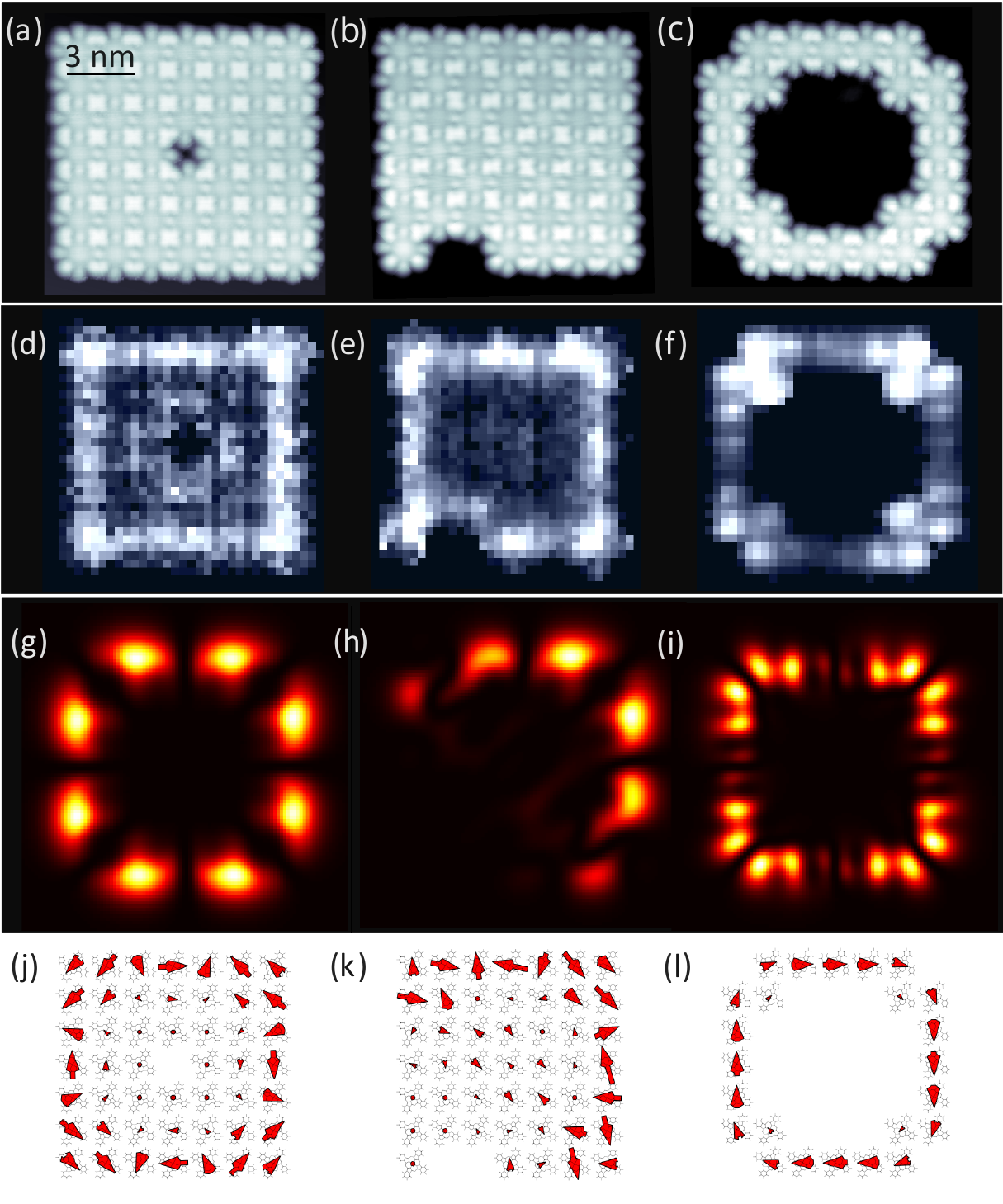}
 \caption{\label{figure4} 
STM topographies ($I$ = 5 pA, $V$ = -2.5 V) of (a) a $7\times 7$ cluster from which the central molecule has been removed, (b) a $7\times 7$ cluster from which two adjacent molecules of the periphery have been removed, and (c) a 24-molecules ring. STML maps (blue, $V = $-2.7 V) of the associated edge states of the clusters (d-e), and of the lowest energy peak of the ring (f). Simulated plasmon-exciton coupling maps (red) of (g-h) the edge-state excitonic modes of the clusters, and of the lowest excitonic mode of the ring (i) obtained from a coherent coupling model of molecular transition dipoles interacting with the plasmonic cavity. (j-l) Corresponding collective dipole configurations for each mode.}
\end{figure}

Tip-manipulation was used to controllably modify a $7 \times 7$ cluster (\Figref{figure4}a,b,c), with the dual objective of probing the sensitivity of the edge state to localised perturbations and testing an alternative strategy for confining excitons along specific paths. In \Figref{figure4}a, the central molecule of the cluster is removed; in \Figref{figure4}b, two edge molecules are removed; and in \Figref{figure4}c, the cluster is reduced to a 24-molecule ring. Experimental STML maps and corresponding theoretical plasmon-exciton coupling maps in \Figref{figure4}d,g,j confirm that the edge state is unaffected by the removal of the central "bulk" molecule. More strikingly, the experimental data in \Figref{figure4}e reveal that the spatial distribution of intensity bypasses the two-molecule edge-vacancy, supporting the potential of these cluster edges to act as efficient excitonic waveguides. Our theoretical simulations, in contrast, show that the edge states are affected by the defect, which splits them into several eigenstates of distinct symmetries and closely spaced energies (the most prominent one is represented in \Figref{figure4}h,k), an effect that the experiment may not capture because of the spectral overlap between these states. Eventually, exciton confinement can be artificially imposed by shaping clusters into single-molecule-wide rings. Experimental and simulated maps of the lowest exciton state (\Figref{figure4}f,i,l) reveal in-phase dipole oscillation around the 24-molecule ring, confirming that coherent coupling is preserved in this case, and providing a complementary strategy to funnel excitons in coherently coupled structures. 


Our findings establish molecular excitonic systems as a new experimental platform where concepts from band theory can be explored at the level of individual quantum states and directly visualized in real space. Looking ahead, the combination of atomic-scale fabrication, molecular self-assembly and excitonic band engineering offers a route toward materials supporting flat bands, correlated excitonic phases, topologically protected transport and novel light-matter collective states. These capabilities could ultimately enable a new generation of molecular architectures in which the flow of optical energy is programmed through band engineering rather than molecular chemistry alone.

\section{Acknowledgements}
G.S. acknowledges funding from the European Research Council (ERC) under the European Union's Horizon 2020 research and innovation program (grant agreement No 101142565), and the Agence Nationale de la Recherche grant ANR-MIDI (ANR-24-CE30-5502). This work of the Interdisciplinary Thematic Institute QMat, as part of the ITI 2021 2028 program of the University of Strasbourg, CNRS and Inserm, was supported by IdEx Unistra (ANR 10 IDEX 0002), and by SFRI STRAT'US project (ANR 20 SFRI 0012) and EUR QMAT ANR-17-EURE-0024 under the framework of the French Investments for the Future Program. T.F. acknowledges support from the Spanish MCIN / AEI / 10.13039/501100011033 (PID2023-146694NB-I00, GRAFIQ). S.N. and T.N. acknowledge the Lumina Quaeruntur fellowship of the Czech Academy of Sciences and QUANTERA project MOLAR with reference PCI2024-153449. Computational resources were provided by the e-INFRA CZ project (ID:90254), supported by the Ministry of Education, Youth and Sports of the Czech Republic.

\clearpage
\setcounter{section}{0}
\setcounter{figure}{0}
\setcounter{table}{0}
\setcounter{equation}{0}

\renewcommand{\thesection}{S\arabic{section}}
\renewcommand{\thefigure}{S\arabic{figure}}
\renewcommand{\thetable}{S\arabic{table}}
\renewcommand{\theequation}{S\arabic{equation}}

\begin{center}
{\large\bfseries SUPPLEMENTARY INFORMATION for\\Topological Frenkel excitons in 2D molecular crystals}
\end{center}
\begin{itemize}

\item[\textbf{S1}] \textbf{Materials and methods} \hfill 17
\begin{itemize}
    \item[\textbf{A}] General experimental method \hfill 17
    \item[\textbf{B}] Acquisition of STML maps \hfill 17
    \item[\textbf{C}] Molecular cluster fabrication \hfill 18
\end{itemize}

\item[\textbf{S2}] \textbf{Adsorption configuration of the clusters on NaCl} \hfill 20

\item[\textbf{S3}] \textbf{Theoretical analysis of electroluminescence spectra and photon maps} \hfill 23
\begin{itemize}
    \item[\textbf{A}] Hamiltonian of an aggregate \hfill 23
    \item[\textbf{B}] TDDFT calculations and effective charges \hfill 24
    \item[\textbf{C}] Plasmon-exciton coupling maps \hfill 25
    \item[\textbf{D}] Averaged STML spectra \hfill 26
    \item[\textbf{E}] Mode selection \hfill 27
\end{itemize}

\item[] \textbf{References} \hfill 35

\end{itemize}
\clearpage
\section{Materials and methods}

\subsection{General experimental method}
STM and STML measurements were carried out in an ultrahigh-vacuum low-temperature Omicron scanning tunneling microscope operating at 5 K and equipped for optical detection of the light emitted from the tunnel junction. The emitted photons were collected through a spectrometer coupled to a CCD camera. Spectra were recorded using a spectral resolution of $\approx$ 4 meV at 1.87 eV.
Ag(111) substrates were prepared by repeated cycles of Ar$^+$ sputtering and annealing. Electrochemically etched silver tips were conditioned in situ by controlled contact with the Ag surface in order to optimize the plasmonic response of the junction.
Approximately 0.5 monolayer of NaCl was deposited onto the clean Ag(111) surface held at room temperature, followed by annealing at 160 $^\circ$C to form square NaCl islands three monolayers thick.
All experiments were performed at 5 K, where atomic diffusion is effectively frozen and the tunnel junction remains highly stable. Under these conditions, the picocavity geometry and the associated localized plasmon modes can be regarded as stationary throughout the measurement.

\subsection{Acquisition of STML maps}

STML maps were acquired in constant-height mode with the feedback loop disabled, ensuring a fixed tip-sample separation and preventing distance-dependent artifacts. For each pixel of the scan, a complete STML spectrum was recorded. Spatial maps corresponding to specific optical transitions were subsequently obtained by integrating the photon counts within selected energy windows of the hyperspectral dataset. This procedure provides, from a single acquisition, a series of directly comparable photon-intensity maps associated with different spectral features.
The longest acquisitions extended up to 16 h. To compensate for residual thermal and mechanical drifts over such durations, an atom-tracking routine was executed before the acquisition of each pixel of the grid. The position of a reference atomic feature was monitored and used to actively correct the tip location, ensuring sub-angstrom stability throughout the measurement.
Photon-intensity maps were finally normalized by the simultaneously recorded tunneling-current map. Since the tunneling current acts as the excitation source and varies with tip position, this normalization compensates for local variations in excitation rate and yields maps reflecting the intrinsic photon-emission efficiency of the junction.

\subsection{Molecular cluster fabrication}

\begin{figure}
  \includegraphics[width=\linewidth]{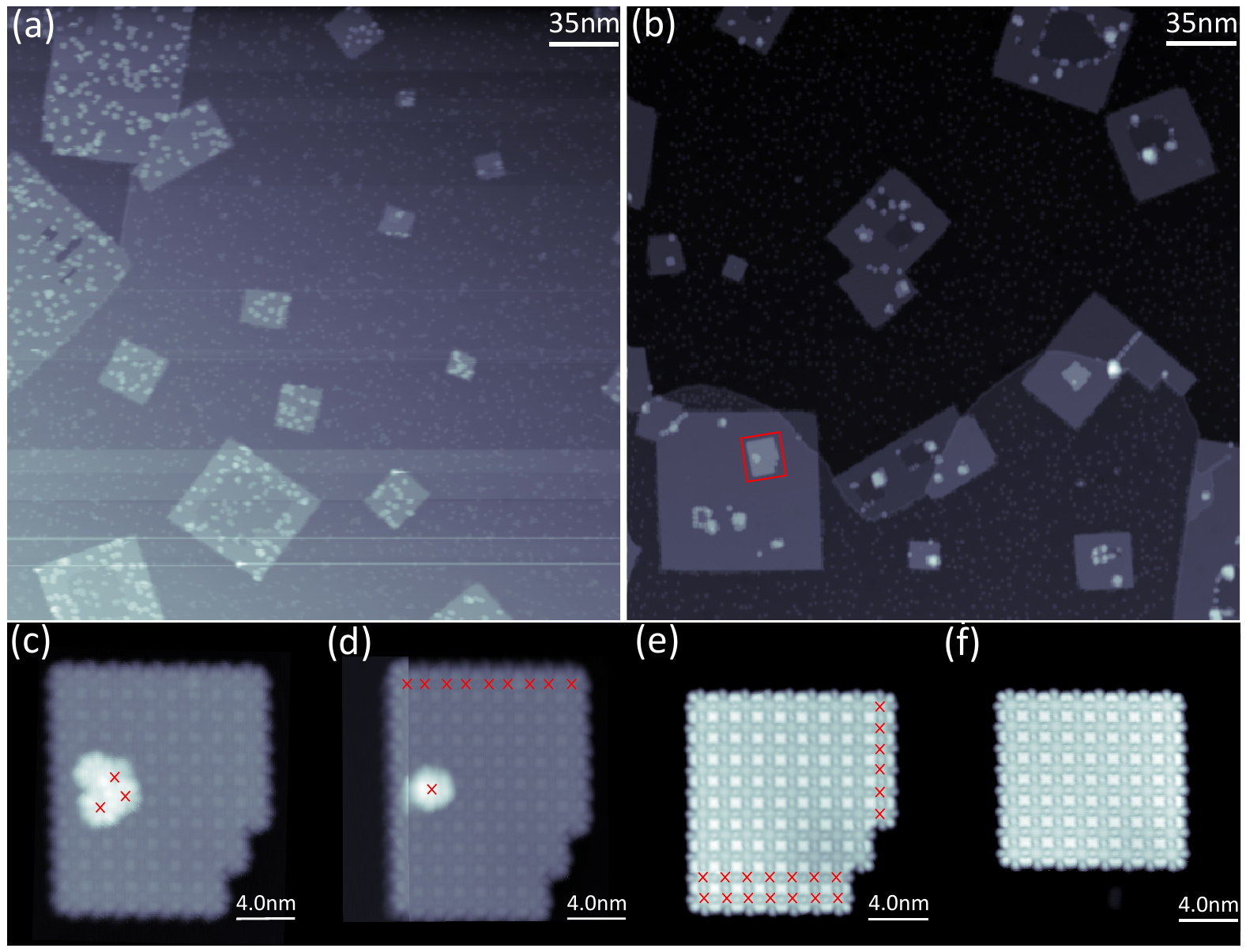}
  \caption{\label{SIfigure1}{(a) Large scale STM image (300 $\times$ 300 nm$^{2}$, $I$ = 10 pA, $V$ = 2 V) of the sample after molecule deposition  onto a sample held at 5 K during deposition. (b) Large scale STM image (300 $\times$ 300 nm$^{2}$, $I$ = 5 pA, $V$ = 2V) of the sample (other area) after a mild thermal annealing of the sample. (c) Constant current ($I$ = 5 pA, $V$ = -2.7 V) STM image of the molecular aggregate framed in red in (b). (c) $\xrightarrow[]{}$ (f) STM images (20 $\times$ 20 nm$^{2}$, $I$ = 5 pA, $V$ = -2.7 V) showing the fabrication of an 8 $\times$ 8 molecular aggregate by tip manipulation. The molecules which were progressively removed from the cluster are marked by a red cross in the preceding image.}}
\end{figure}

The realization of the 2D molecular clusters relies on a succession of steps: (i) Sublimation of ZnPc molecules onto a NaCl-covered Ag(111) sample at 5 K, yielding submonolayer coverage. At that stage the molecules are randomly scattered on the surface (\Figref{SIfigure1}a). (ii) Short thermal annealing of the sample to provide sufficient kinetic energy for the molecules to self-assemble in compact 2D clusters on NaCl (\Figref{SIfigure1}b,c). (iii) ''Cleaning'' the clusters from the few molecules that may be adsorbed on top (\Figref{SIfigure1}d,e), and (iv) ''trimming'' to form clusters of the desired size and shape (\Figref{SIfigure1}f). The cleaning and trimming steps are based on a succession of controlled STM tip manipulations.  In detail, the tip is slowly approached to the molecule to be removed until contact is reached. As the interaction with the metallic tip is stronger than the one with NaCl (and with the first layer of molecules), the targeted molecule is transferred (with a reasonable success rate), on the shaft of the tip. This molecule is then released by a small tip indentation in the metal surface. The procedure is performed again until the desired structure is formed.

\clearpage
\section{Adsorption configuration of the clusters on NaCl}

\begin{figure}
  \includegraphics[scale=1.2]{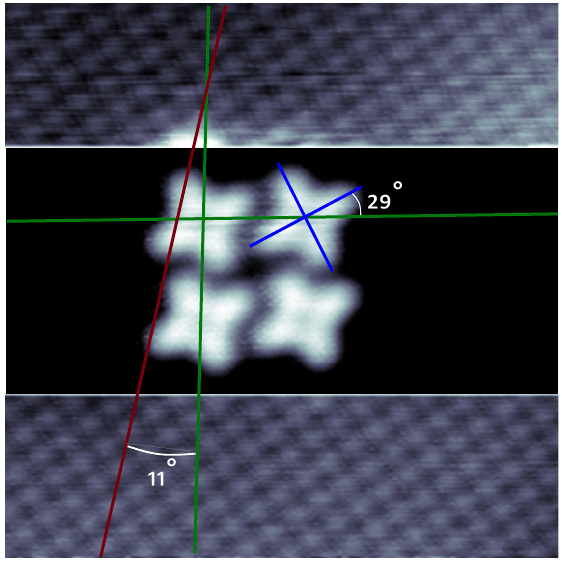}
  \caption{\label{SIfigure2} {STM image (8.0 $\times$ 8.0 nm$^{2}$) showing the epitaxial organisation of ZnPc molecules within a 2 $\times$ 2 molecular cluster on top of a NaCl island. The first part of the image was acquired in constant current mode with an initial setpoint of ($I$ = 150\,pA, $V$ = 100\,mV) in order to image the NaCl island with atomic resolution. Upon reaching the molecular aggregate, the setpoint was switched to ($I$ = 5\,pA, $V$ = 600\,mV), revealing an in-gap image of the molecular cluster. After leaving the molecular aggregate, the setpoint was switched back to ($I$ = 150\,pA, $V$ = 100\,mV). The red line drawn along one row of Na-Cl atoms and the green line drawn along one row of ZnPc molecules allow for the measurement of an  $\approx 11^\circ$ angle between the highest linear atomic density axis of the NaCl lattice and the main axis of the molecular lattice.}}
\end{figure}

The exact adsorption geometry of a 2 $\times$ 2 ZnPc cluster with respect to the NaCl lattice was determined experimentally. We launch the acquisition of the STM image displayed in \Figref{SIfigure2} with a high current (150 pA) and a low sample voltage (100 mV). Under these conditions, the atomic positions of the Cl atoms in the top NaCl layer are clearly resolved. Shortly before the ZnPc cluster entered the scan area, the tunneling current was reduced to 5 pA and the voltage increased to 600 mV, thereby increasing the tip--sample separation and enabling the safe imaging of the molecules. Once the molecules were no longer within the scan range, the current was restored to 150 pA, and the Cl atoms are resolved again.

This imaging procedure provides a direct experimental determination of the adsorption configuration of the self-assembled molecules on the NaCl surface. It reveals that the ZnPc molecules are adsorbed on top of Cl sites and that the principal axis of the molecular cluster is rotated by $ \approx 11^\circ$ with respect to the dense-packed direction of the NaCl lattice. The relative orientations and positions of the individual molecules within the cluster can also be determined with high precision. These experimental parameters were subsequently used to set the precise molecule-molecule distance and  respective orientation used in the model presented in the manuscript.

\begin{figure}
  \includegraphics[width=\linewidth]{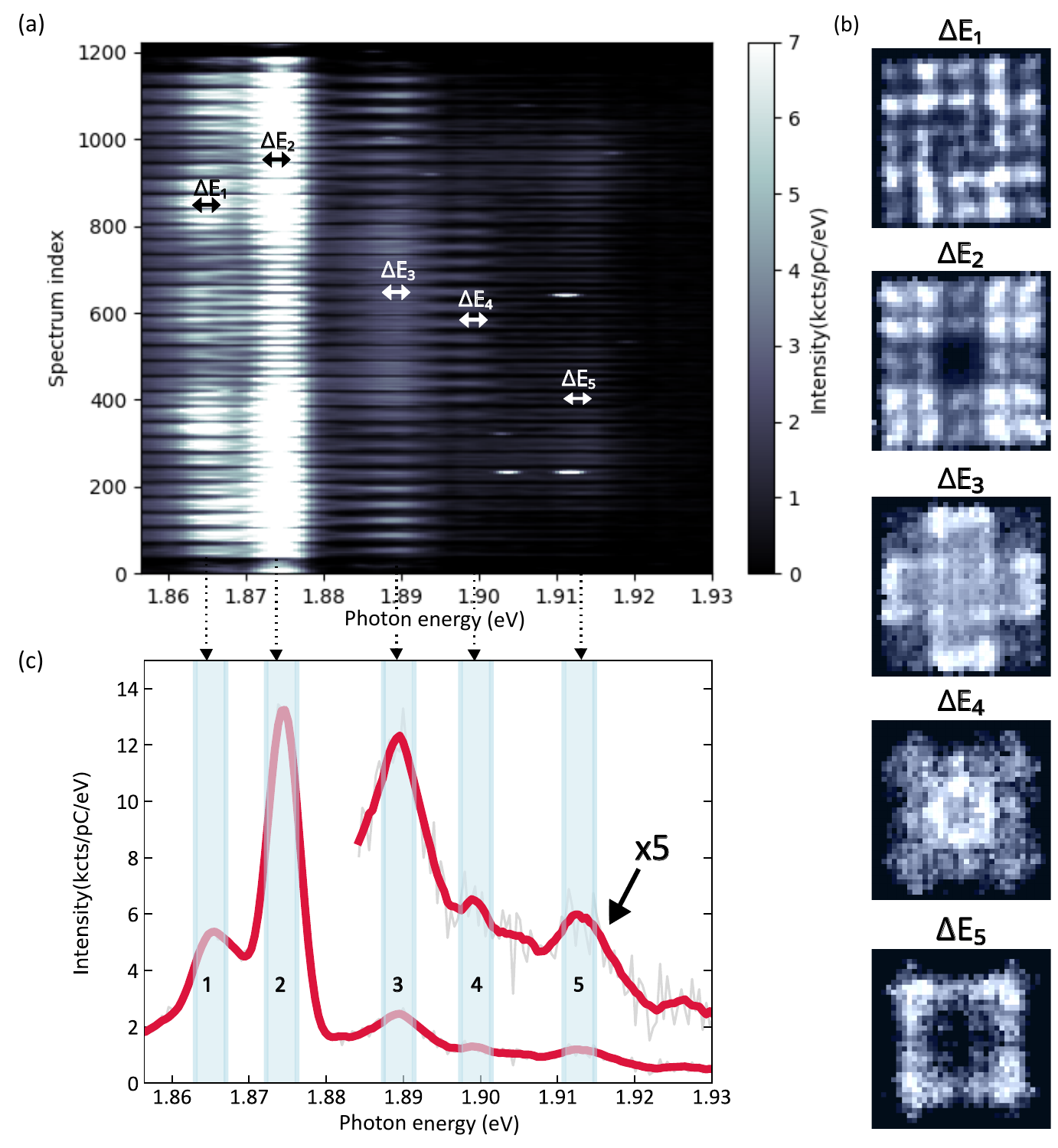}
  \caption{\label{SIfigure3}{(a) Intensity map obtained by compiling the 1225 STML spectra acquired over a 5$\times$5 molecular aggregate. The energy windows corresponding to the five identified feature are denoted by $\Delta \textnormal{E}_{1\xrightarrow{}5}$}. (b) Hyperspectral fluorescence maps (8.3 $\times$ 8.3 nm$^{2}$) obtained by integrating the same 1225 STML spectra over the energy windows given in (a). (c) Averaged spectrum (same as the one in Figure 1c in the main manuscript.)}
\end{figure}
\clearpage
\newpage
\section{Theoretical analysis of electroluminescence spectra and photon maps}
In the main text, we discuss electroluminescence spectra and spectrally-resolved photon maps and assign to each spectral peak a dominantly contributing excitonic mode of the cluster. Here, we discuss the theoretical calculations which support our selection of the modes.  
\subsection{Hamiltonian of an aggregate}	
\begin{figure}
    \centering
    \includegraphics[width=\linewidth]{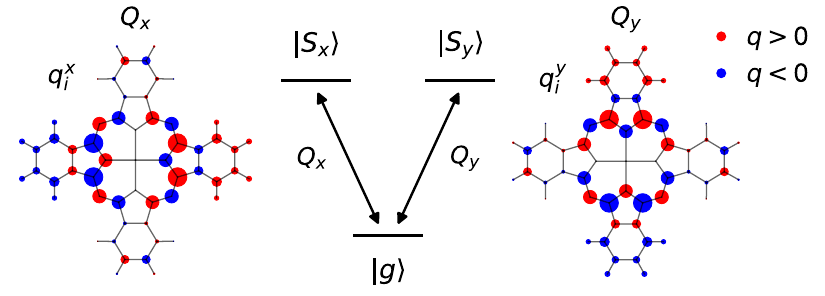}
    \caption{Schematic representation of the monomer Hamiltonian composed of the ground state $|g\rangle$ and a pair of degenerate excited states $|S_x\rangle$ and $|S_y\rangle$. We also mark the associated transitions denoted $Q_x$ and $Q_y$, respectively, and show their corresponding transition charge densities represented using the equivalent charges $q_i^{\alpha}$ ($\alpha\in\{x,y\}$) positioned on every atom $i$ of the molecule. Red (blue) dots denote positive (negative) charges.}
    \label{fig:sketch_transitions}
\end{figure}

The aggregates are squares of $N\times N$ ZnPc molecules which are arranged in square lattice and structures built by controllably removing molecules from them. For each molecule $m$ we consider three electronic states: the ground state $\ket{g_{m}}$, and two degenerate singlet excited states $\ket{S_{\alpha}^{m}}$ with $\alpha = x, y$. 
We restrict the Hilbert space to the single-excitation manifold. The corresponding basis states of the aggregate are
\begin{equation}
\ket{m,\alpha}=\ket{S_\alpha^{m}}\prod_{n\neq m}\ket{g_n},
\end{equation}
while the aggregate ground state is
\begin{equation}
\ket{g_{\mathrm{agg}}}=\prod_{m=1}^{N_{\rm mol}}\ket{g_m}.
\end{equation}
Here $N_{\rm mol}$ is the number of molecules in the aggregate. For further convenience, we offset the energies of the molecular excitations such that zero energy corresponds to the excitation energy of the isolated monomer. The Hamiltonian of the system then reduces to the inter-molecular terms mediated by the Coulomb coupling between the
$\alpha$-polarized transition of molecule $m$ and the $\beta$-polarized
transition of molecule $n$:
\begin{equation}
	\hat H= \sum_{m\neq n}\sum_{\alpha,\beta\in\{x,y\}}V_{mn}^{\alpha\beta}\ket{n,\beta}\bra{m,\alpha},
	\label{eqs:hamiltonian}
\end{equation}
with
\begin{equation}
	V_{mn}^{\alpha\beta}
	= \sum_{i=1}^{N_{\rm at}}\sum_{j=1}^{N_{\rm at}}
	\frac{q_i^{\alpha}q_j^{\beta}}{\left|(\mathbf R_m+\mathbf r_i)-
		(\mathbf R_n+\mathbf r_j)\right|},
	\qquad m\neq n.
	\label{eq:exciton_coupling}
\end{equation}
Here $q_i^{\alpha}$ are point transition
charges, each localized on atom $i$ of the $N_{\rm at}$-atomic molecule, shown in Fig.\,\ref{fig:sketch_transitions}. These charges represent the transition charge density associated with the $\alpha$-polarized
transition of molecule $m$:
\begin{equation}
\rho_m^{\alpha}(\mathbf r)\approx\sum_{i=1}^{N_{\rm at}}q_i^{\alpha}\delta\left(
\mathbf r-\mathbf R_m-\mathbf r_i\right),
\end{equation}
where $\mathbf R_m$ denotes the center of molecule $m$, while $\mathbf r_i$
is the position of charge $i$ relative to the molecular center. The molecules are distributed on a square lattice with lattice parameter $a_0=1.45$\,nm. Each molecular monomer is rotated within the unit cell by an angle of $\phi=29^\circ$, as shown in Fig.\,\ref{SIfigure2}.
\subsection{TDDFT calculations and effective charges}

The intermolecular interactions are evaluated by representing the molecular transition densities $\rho_\alpha\approx \sum_i q_i^\alpha\delta({\bf r}-{\bf r}_i)$ ($\alpha\in\{x,y\}$) with sets of atom-centered equivalent charges $q_i^\alpha$ positioned at locations ${\bf r}_i$ of atom $i$. The transition densities $\rho_\alpha$ are first obtained from linear-response time-dependent density-functional theory calculations carried out using Gaussian 16, revision C.01 \cite{g16}. The calculations employ the B3LYP hybrid functional \cite{Becke1993, Lee1988, Becke1993a} together with the Pople 6-31G* basis set. The ground-state molecular geometry is optimized at the same level of theory while imposing D$_{4\rm h}$ symmetry.

The resulting transition densities are processed with Multiwfn 3.8 \cite{Lu2012} and represented by charges fitted to their electrostatic potentials using the grid-based CHELPG procedure. The fitting is performed through the implementation of the CHELPG method in the LIBRETA library \cite{Zhang2021b}, interfaced with Multiwfn. The atom-centered charges are determined by fitting their electrostatic potential to that generated by the full three-dimensional transition density at grid points surrounding the molecule. We employ a grid extending up to $20$\,Bohr from the molecular atomic centers with a spacing of $0.8$\,Bohr. Grid points closer to the atoms than the minimum distances specified by the LIBRETA library are excluded. We use the default values of 1.450\,{\AA}, 1.5\,{\AA}, and 1.7\,{\AA} for H, C, and N, respectively, and an atomic radius of 1.1513\,{\AA} for Zn.

\subsection{Plasmon-exciton coupling maps}
Here we describe how we obtain the plasmon-exciton coupling maps and evaluate polarization textures of the eigenstates. The outcoupling probability of an exciton $\ket{\Psi^{(k)}}$ into the far field through the interaction with the tip plasmon is considered to be proportional to the square modulus $|G_k|^2$ of the plasmon-exciton coupling parameter defined as an overlap integral of the excitonic transition charge density $\rho^{(k)}$ and the plasmonic quasi-static potential $\phi_{\rm pl}$:
\begin{align}
    G_k({\bf r}_{\rm tip})=\int \rho^k({\bf r})\phi_{\rm pl}({\bf r}-{\bf r}_{\rm tip})\,{\rm d}^3{\bf r},
\end{align}
where ${\bf r}_{\rm tip}=(x_{\rm tip},y_{\rm tip}, z_{\rm const})$ is the position of the tip in the scanning plane (assuming a constant height $z_{\rm tip}=z_{\rm const}$). 

To obtain the transition density $\rho^k$ of the exciton we diagonalize the Hamiltonian in Eq.\,\eqref{eqs:hamiltonian}. The $k$-th eigenstate of $\hat H$ can then be expanded in the localized molecular basis as
\begin{equation}
	\ket{\Psi^{(k)}}=\sum_{m=1}^{N_{\rm mol}}\sum_{\alpha\in\{x,y\}}c_{\alpha, m}^{(k)}\ket{m,\alpha}.
	\label{eq:eigenstate}
\end{equation}
Using this expansion, we construct transition-charge distribution for each eigenstate using the approximation employing the transition charges of the monomer.
Namely, the transition charge associated with atom $i$ of molecule $m$ in the
eigenstate $\ket{\Psi^{(k)}}$ is obtained as
\begin{equation}q_{m,i}^{k}=c_{x,m}^{(k)}q_i^{x}+c_{y, m}^{(k)}q_i^{y}.
\label{eq:effective_charge}
\end{equation}
We assume that the molecules are planar and placed in the $z=0$ plane. These transition charges $q_{m,i}^{k}$ are then also used to define the transition dipoles localized on each molecule, which we visualize using molecule-centered red arrows in the main text. 

The quasi-static potential of the tip plasmon is modeled using two point charges of opposite sign and equal magnitude located at ${\bf R}_1(x_{\rm tip},y_{\rm tip})=(x_{\rm tip},y_{\rm tip},z_1)$ and ${\bf R}_2(x_{\rm tip},y_{\rm tip})=(x_{\rm tip},y_{\rm tip},z_2)$, with $z_1=1\,{\rm nm}$ and $z_2=-2\,{\rm nm}$.
The position dependent coupling between the tip and an aggregate in its excitonic eigenstate
$\ket{\Psi^{(k)}}$ is then evaluated for a tip position $(x_{\rm tip},y_{\rm tip})$ as
\begin{equation}
	G_k(x_{\rm tip},y_{\rm tip})=\sum_{m=1}^{N}\sum_{i=1}^{N_q}q_{m,i}^{(k)}\left[\frac{1}{\left|\mathbf R_1(x_{\rm tip},y_{\rm tip})-(\mathbf R_m+\mathbf r_i)\right|}-\frac{1}{\left|\mathbf R_2(x_{\rm tip},y_{\rm tip})-(\mathbf R_m+\mathbf r_i)\right|}\right].
	\label{eq:g}
\end{equation}
The corresponding spatially resolved plasmon-exciton coupling map is defined as
\begin{equation}
	\mathcal I^{(k)}(x_{\rm tip},y_{\rm tip})=\left|G_{k}(x_{\rm tip},y_{\rm tip})\right|^2.
	\label{eq:Intensity2}
\end{equation}
Thus, $\mathcal I^{(k)}(x_{\rm tip},y_{\rm tip})$ provides a spatial map of the coupling of
eigenstate $k$ to the tip plasmon as the tip is scanned across the molecular
aggregate. 

\subsection{Averaged STML spectra}
The aggregates containing $N\times N$ molecules support $2\times N\times N$ singly excited excitons. On the other hand, the experimental spectra contain only a small number of clearly identifiable spectral peaks. It is therefore important to establish a strategy that allows us to assign the spectral peaks to the most relevant underlying excitonic modes that contribute to it. To that end we approximate the experimental averaged STML spectra as
\begin{equation}
S_{T}(\omega, T_{\rm eff})=\sum_iP_i(T_{\mathrm{eff}})\Gamma_i^{\mathrm{rad}}
\frac{1}{\pi}
\frac{\gamma_i/2}
{
\left(\omega-E_i\right)^2
+\left(\gamma_i/2\right)^2},\label{eq:temperature_spectrum}
\end{equation}
where $\Gamma_i^{\mathrm{rad}}=
\alpha_{\mathrm{pt}} W^{(i)}$, with $W^{(i)}=\sum_{p,q}\mathcal I^{(i)}(x_p,y_q)$ being an averaged tip-dependent outcoupling rate of the exciton $i$, and $\gamma_i=6~\mathrm{meV}$ is a broadening parameter chosen to approximate the experimentally detected spectral width of the states accounting for broadening due to both decay and pure dephasing processes, including vibrationally driven internal conversion. Finally, 
\begin{equation}
    P_i(T_{\mathrm{eff}})=
    \frac{\exp[-(E_i-E_0)/(k_{\mathrm B}T_{\mathrm{eff}})]}
    {\sum_j \exp[-(E_j-E_0)/(k_{\mathrm B}T_{\mathrm{eff}})]},
\end{equation}
is a population of state $i$ approximated using a Boltzmann distribution for an effective temperature $T_{\mathrm{eff}}$. Here $E_0$ is the energy of the lowest excited state, $E_i$ is the energy of state $|\Psi^{(i)}\rangle$, and $k_{\rm B}$ is the Boltzmann constant. The effective temperature $T_{\mathrm{eff}}\approx 100$\,K results in populations that generate a spectrum with relative peak intensities comparable to the experiment. We note that our approach, assuming thermally distributed populations, is approximate and cannot fully account for the vibronically driven dynamics of pumping, radiative and non-radiative decay, and internal conversion that establish the non-equilibrium populations. The full treatment of these properties is beyond the scope of the present theory. 

\subsection{Mode selection}
\begin{figure}[t]
    \centering
    \includegraphics[
        width=\textwidth,
        keepaspectratio]{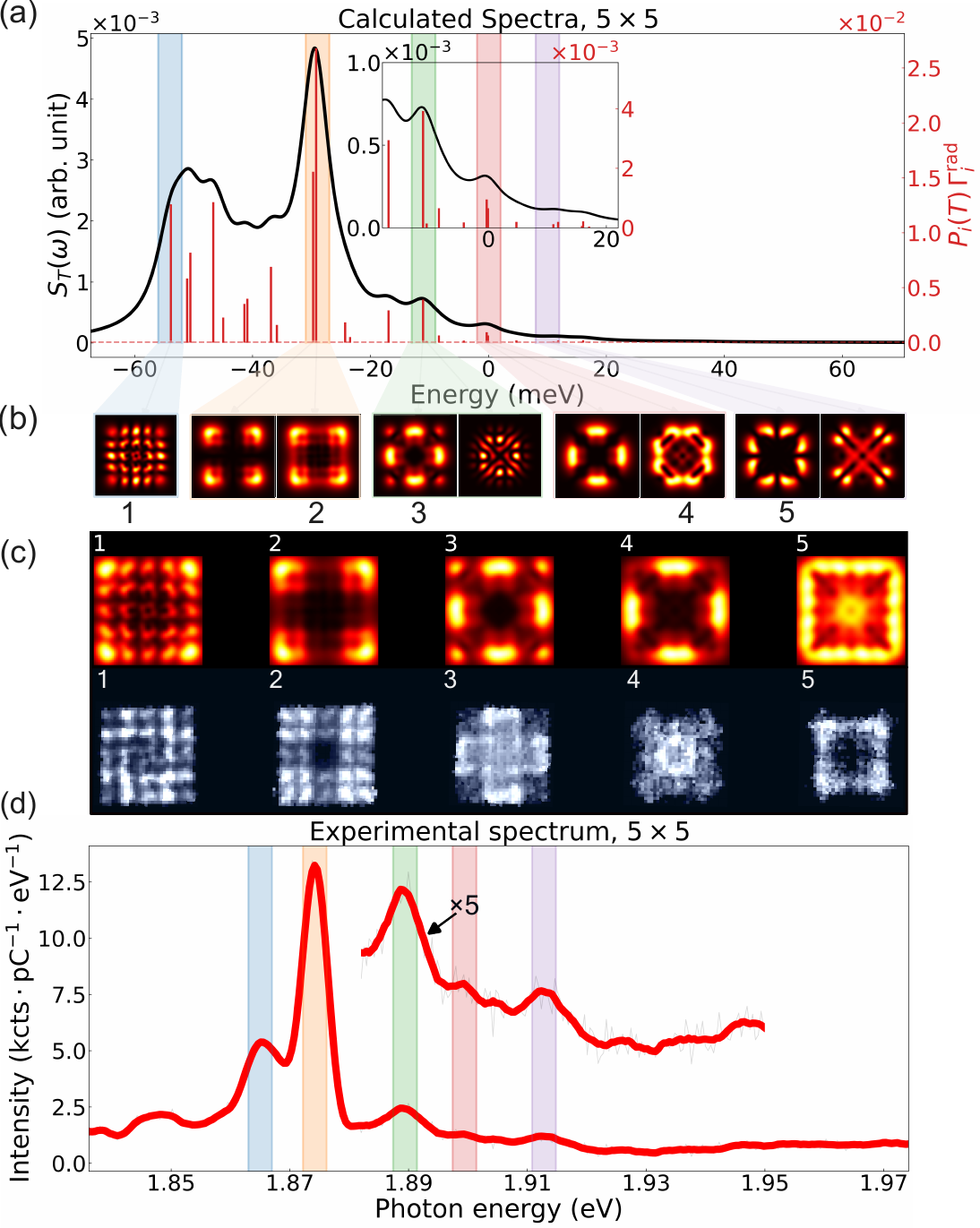}
    \caption{
   {$5\times5$ aggregate.} Effective temperature $T_{\rm {eff}} = 100{\rm{K}}$. }
    \label{fig:SI_Fig_5_5}
\end{figure}

To identify the modes predominantly associated with the experimental emission maps, we first calculate the theoretical spectrum using Eq.~({\ref{eq:temperature_spectrum}}), as shown in Fig.~{\ref{fig:SI_Fig_5_5}}a using an example of a $5\times 5$ cluster. In the theoretically calculated spectrum, we can clearly identify one dominant spectral peak accompanied by a broader, less prominent, peak at lower energy and a series of weaker peaks at higher energy. The overall pattern observed in the spectrum closely resembles that of the experimental spectrum in Fig.\,\ref{fig:SI_Fig_5_5}d, although the relative peak energies differ slightly, likely due to the approximations employed in the theoretical model. 
We next select spectral windows of width $4.0$\,meV and vary their central energies, calculating the spectrally filtered photon maps for each. Using this procedure, we identify five relevant spectral windows, showing a good match with the experimental maps. We show the side-by-side comparison of the theory maps (top row) and experimental maps (bottom row) in Fig.\,\ref{fig:SI_Fig_5_5}c. 
We then identify the most prominent modes that contribute to the spectrum. In Fig.\,\ref{fig:SI_Fig_5_5}a we show their spectral weights $A_i=P_i\Gamma_i^{\mathrm{rad}}$ as red stems at the corresponding energies.
We plot in Fig.\,\ref{fig:SI_Fig_5_5}b the plasmon-exciton coupling maps of the modes contributing to the selected spectral windows (the sum of maps in the case of degenerate modes). 

Based on the resemblance of the simulated and experimental averaged STML spectra, the similarity of the spectrally filtered maps obtained from theory and the experiment, and the theoretical analysis of the underlying mode contributions, we finally select the representative modes that we present in the main text to explain the origin of the experimentally obtained patterns. 

\begin{figure}[t]
    \centering
    \includegraphics[
        width=\textwidth,
        keepaspectratio]{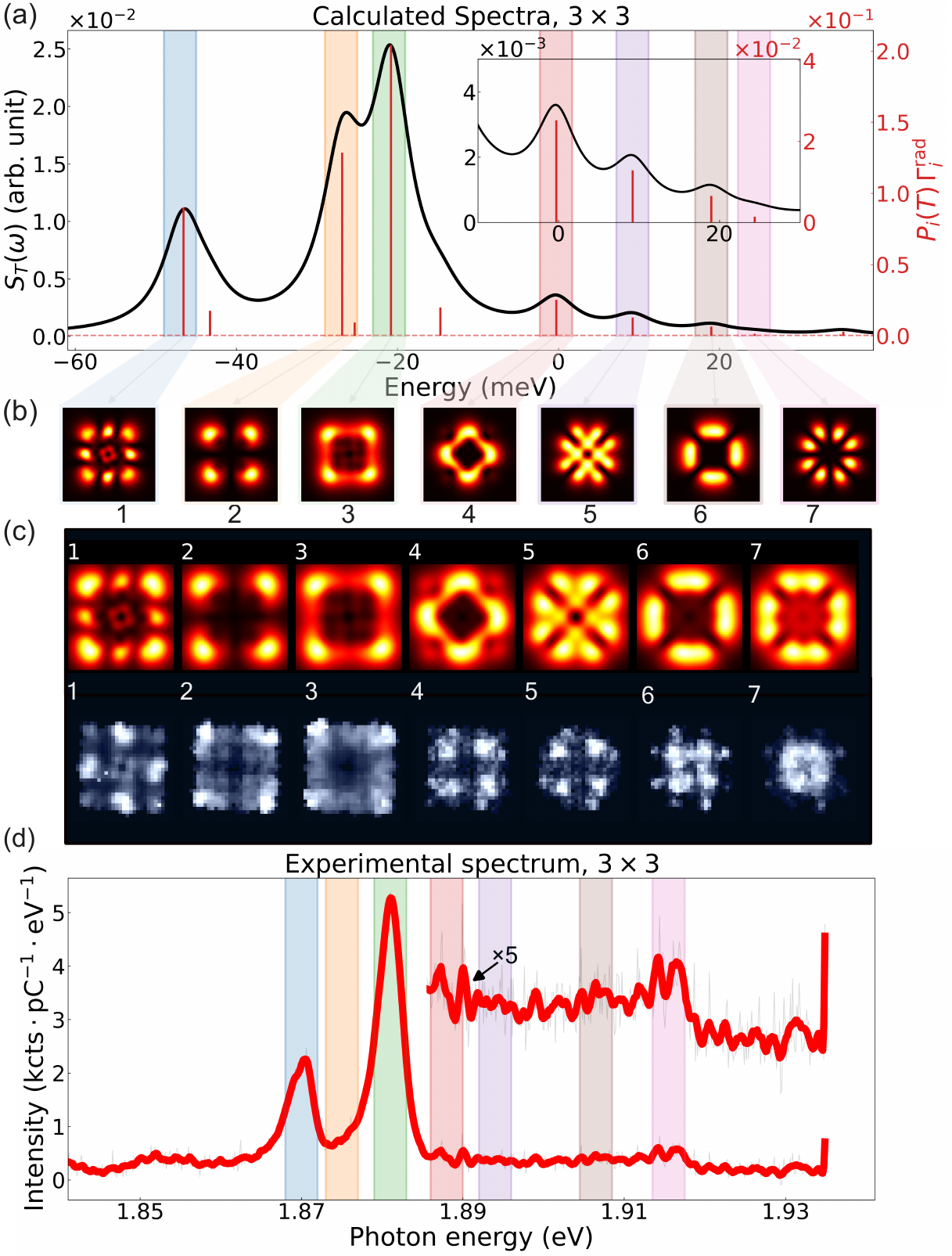}
    \caption{{$3\times3$ aggregate.} Effective temperature $T_{\rm {eff}} = 150{\rm{K}}$. }
    \label{fig:SI_Fig_3_3}
\end{figure}

We perform a similar analysis for the other clusters and show the results in Fig.\,\ref{fig:SI_Fig_3_3} to Fig.\,\ref{fig:SI_Fig_8_8}. Overall, we find a good correspondence between the theoretically calculated results and experiment. 

All figures Fig.\,\ref{fig:SI_Fig_3_3} to Fig.\,\ref{fig:SI_Fig_8_8} use the same parameters except for the effective temperature, which is higher for the $3\times 3$ cluster, $T_{\rm eff}=150$\,K, than for all remaining cases, where $T_{\rm eff}=100$\,K. The figures are arranged as follows: (a) Theoretical spectrum $S_{\rm T}(\omega)$ (black curve). The red stems indicate the spectral weights $P_i \Gamma^{\rm rad}_i$ of selected excitonic modes. (b) Plasmon-exciton coupling maps $\mathcal{I}^{(i)}$ for the spectrally prominent modes. The labels $1$-$5$ ($1$-$7$ for the $3\times 3$ cluster) identify the modes associated with the spectrally filtered maps in (c). (c) Top row: spectrally filtered theoretical maps obtained by averaging over the corresponding spectral regions indicated by the colored regions in (a). Bottom row: experimental photon maps obtained from the colored spectral regions indicated in (d). (d) Experimental emission intensity as a function of photon energy.

\begin{figure}[t]
    \centering
    \includegraphics[
        width=\textwidth,
        keepaspectratio]{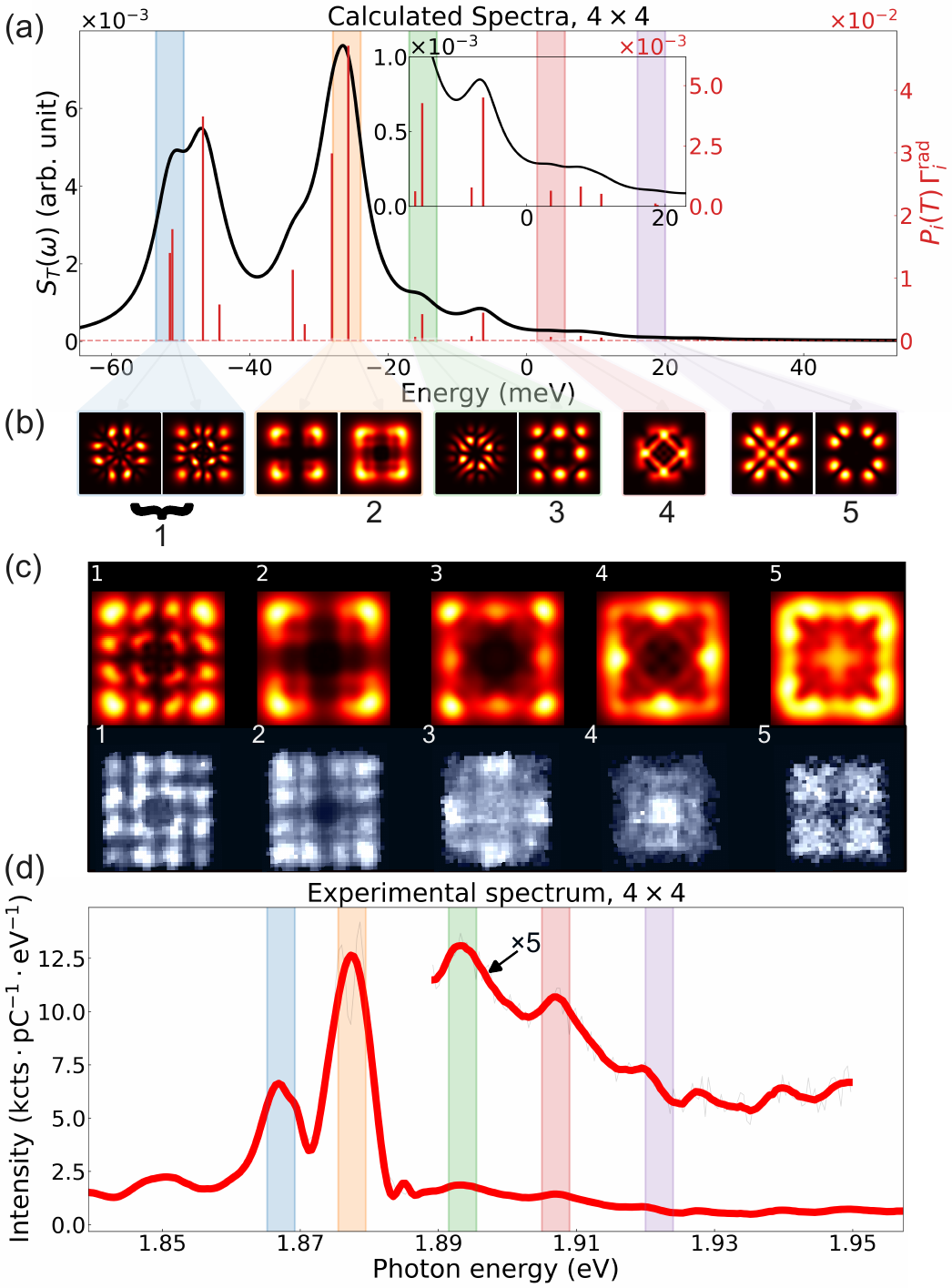}
    \caption{{$4\times 4$ aggregate.} Effective temperature $T_{\rm {eff}} = 100{\rm{K}}$.}
    \label{fig:SI_Fig_4_4}
\end{figure}

\begin{figure}[t]
    \centering
    \includegraphics[
        width=\textwidth,
        keepaspectratio]{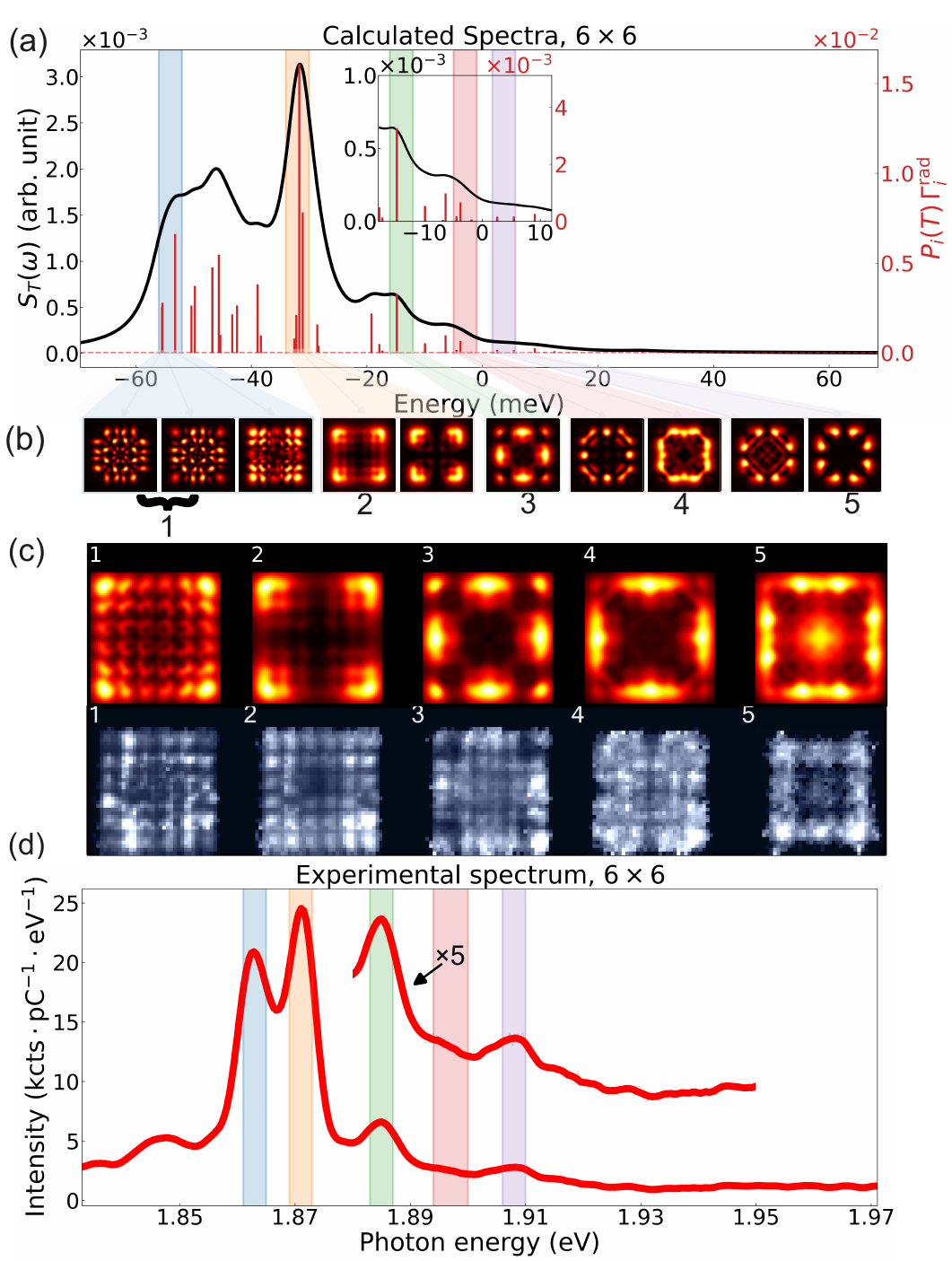}
    \caption{
    {$6\times6$ aggregate.} Effective temperature $T_{\rm {eff}} = 100{\rm{K}}$.
    }
    \label{fig:SI_Fig_6_6}
\end{figure}
\begin{figure}[t]
    \centering
    \includegraphics[
        width=\textwidth,
        keepaspectratio]{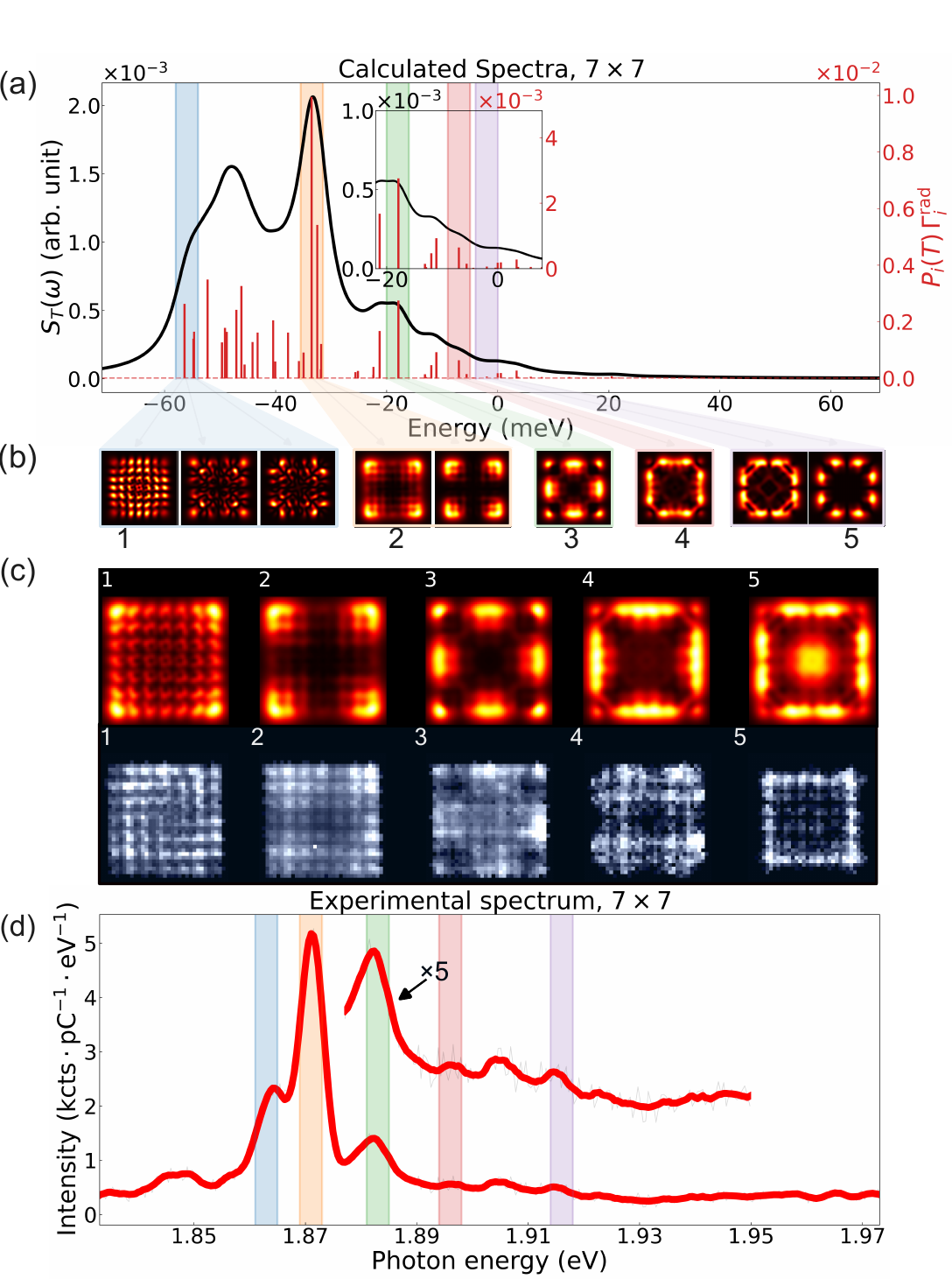}
    \caption{
    {$7\times7$ aggregate.} Effective temperature $T_{\rm {eff}} = 100{\rm{K}}$.
    }
    \label{fig:SI_Fig_7_7}
\end{figure}
\begin{figure}[t]
    \centering
    \includegraphics[
        width=\textwidth,
        keepaspectratio]{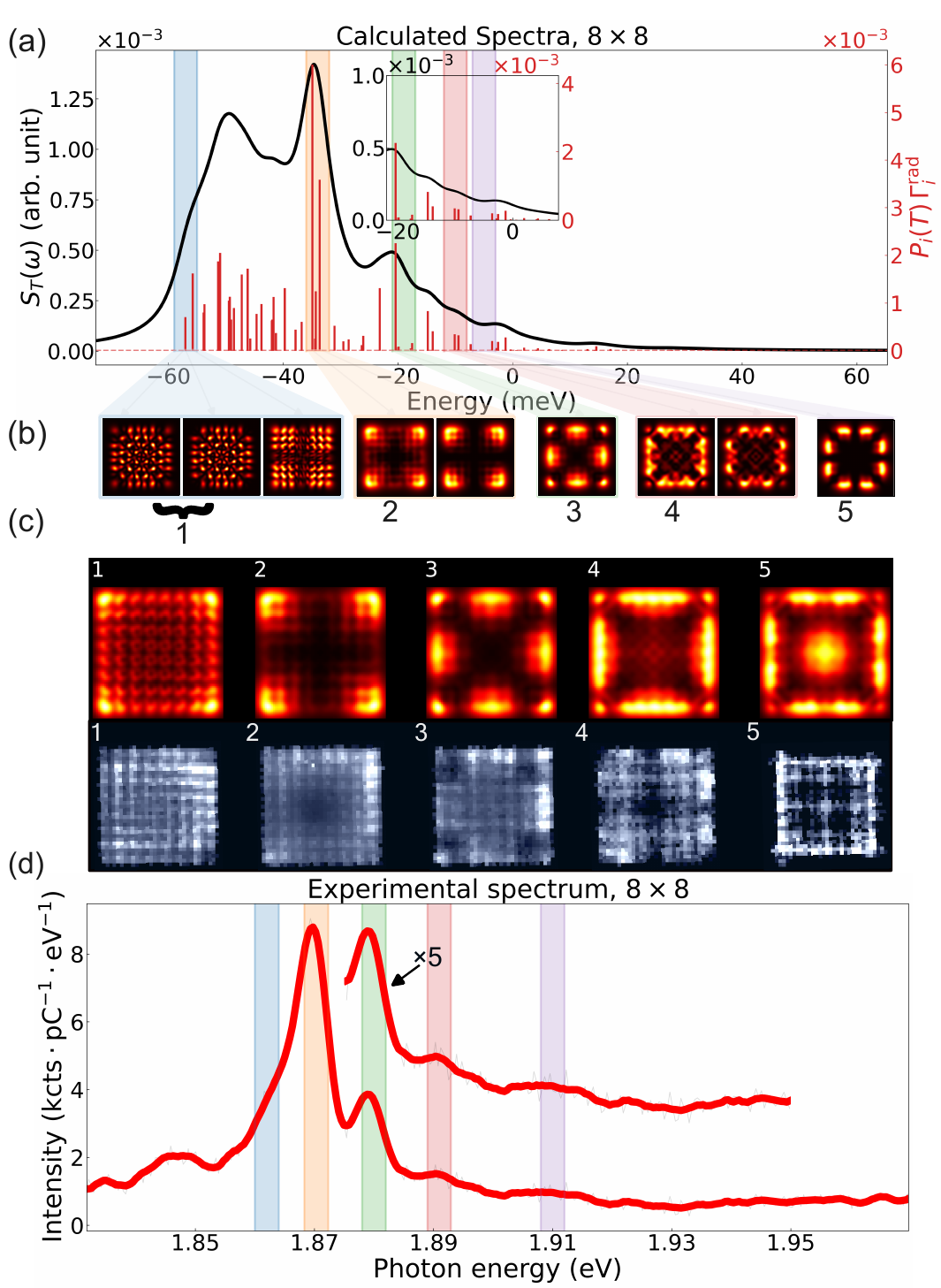}
    \caption{
    {$8\times8$ aggregate.} Effective temperature $T_{\rm {eff}} = 100{\rm{K}}$.
    }
    \label{fig:SI_Fig_8_8}
\end{figure}
\clearpage
%

\end{document}